\documentclass[floatfix,aps,twocolumn,prx,superscriptaddress]{revtex4-2}
\usepackage{graphicx}
\usepackage{dcolumn}
\usepackage{bm}
\usepackage[T1]{fontenc}
\usepackage{textcomp}
\usepackage{amsmath}
\usepackage{mathrsfs}
\usepackage{graphicx}
\usepackage{amssymb}
\usepackage{units}
\usepackage{xcolor}
\usepackage{float}
\usepackage[colorinlistoftodos,prependcaption,textsize=tiny]{todonotes}

\renewcommand{\selectlanguage}[1]{}

\begin{document}

\title{Fermionization and collective excitations  of 1D polariton lattices}

\author{Johannes Knörzer}
\affiliation{Institute for Theoretical Studies, ETH Zurich, 8092 Zurich, Switzerland}
\author{Rafa\l{} O\l{}dziejewski}
\affiliation{Centre for Quantum Optical Technologies, Centre of New Technologies, University of Warsaw, S. Banacha 2c, 02-097 Warsaw, Poland}
\author{Puneet A. Murthy}
\affiliation{Institute for Quantum Electronics,
ETH Zurich, Zurich, Switzerland}
\author{Ivan Amelio}
\affiliation{Center for Nonlinear Phenomena and Complex Systems,
Universit{\'e} Libre de Bruxelles, CP 231, Campus Plaine, B-1050 Brussels, Belgium}

\begin{abstract}
We theoretically demonstrate that the hallmarks of correlation and fermionization in a one-dimensional exciton-polaritons gas can be observed with state-of-the-art technology.
Our system consists of a chain of excitonic quantum dots coupled to a photonic waveguide, with a low filling of polaritons.
We analytically identify the Tonks-Girardeau, Tavis-Cummings and mean-field limits and relate them to different regimes of the excitonic anharmonicity and photonic bandwidth.
Using matrix-product states, we numerically calculate the ground-state energies, correlation functions and dynamic structure factor of the system.
In particular, the latter has a finite weight in the Lieb-Liniger hole branch, and the density-density correlator displays Friedel-like oscillations for realistic parameters, which reveal the onset of fermionization close to the Tonks-Girardeau regime.
Our work encourages future experiments aimed at observing, for the first time and in spite of the moderate excitonic anharmonicity, strongly correlated exciton-polariton physics. 
\end{abstract}

\date{\today}

\maketitle

\section{Introduction}

Reducing the dimensionality of many-body systems fundamentally affects their properties.
One of the most prominent examples is the one-dimensional Tonks-Girardeau (TG) gas of bosons, where the presence of strong inter-particle interactions leads to fermionization~\cite{tonks_complete_1936,girardeau_relationship_2004}.
The TG gas is a limiting case of the underlying Lieb-Liniger (LL) model, which describes a one-dimensional system of repulsively interacting bosons~\cite{lieb_exact_1963}.
In contrast to higher dimensions, the excitation spectrum of the LL model contains two branches in the excitation spectrum: particle and hole excitations, the latter being a unique feature of one-dimensional physics.
As the interaction parameter is tuned from the weak to the strong-coupling regime, the spectrum gradually changes its character from particle to hole-like~\cite{lieb_exact_1963-1}.
Observing this transition in an experiment provides a clear fingerprint of strong interactions in a one-dimensional system~\cite{meinert_probing_2015}.

The paradigmatic example of the 1D Bose gas suggests that dimensional reduction can be a strategy 
to achieve strongly correlated photon 
fluids, which is one main
goal of research in quantum optics,
even in the presence of moderate interactions.
In particular, we focus our attention here on exciton-polaritons~\cite{deng2010,carusotto_quantum_2013}, but our results should also apply to other kinds of polaritons, such as Rydberg polaritons~\cite{gorshkov2011,Peyronel2012}.
Exciton-polaritons have emerged in the last  two decades as an appealing platform to study superfluidity~\cite{Amo2009}, nonlinear physics~\cite{Baas2004,Amo2016}, analogue gravity~\cite{nguyen2016,Jacquet2020}, and non-equilibrium quasi-condensates~\cite{Kasprzak2006,Byrnes2014,Fontaine2022}.
The photon endows the polariton with a light mass, contrasting the detrimental effects of temperature and disorder, and it allows for efficient optical driving and imaging.
However, interactions between polaritons, which are inherited by their matter component, are typically small, such that in two dimensions a mean-field description is accurate.
Second-order correlation measurements reveal up to a modest 5\% of anti-bunching~\cite{Delteil2019}.

In spite of these difficulties, there has been a lot of recent theoretical and experimental work aimed at pushing polaritons into the strongly correlated regime.
Dipolaritons, obtained by hybridizing the photon with an interlayer exciton possessing a permanent dipole moment~\cite{Cristofolini2012,Rosenberg2018,Togan2018}, and polaron-polaritons, where the matter excitation role is played by a charged trion~\cite{tan2020}, have provided an enhancement of one order of magnitude of the interaction over linewidth ratio.
Rydberg excitons in Cu$_2$O compounds offer yet another option to pursue in this direction~\cite{Walther2018,Orfanakis2022}. 
Quantum confinement of excitons to  1D~\cite{Thureja2022} and 0D~\cite{Hu2024, thureja2024electrically} traps  by strongly localized electric fields has recently been achieved.
In many of these advancements, a major role is played by atomically thin semiconductors based on transition-metal dichalcogenides (TMD)~\cite{Mak2022}.

A first hint of the nontrivial role of  dimensionality and kinetic energy in renormalizing the polariton-polariton interaction has been investigated in Ref.~\cite{christensen_microscopic_2022}.
Here, the authors studied two dipolaritons on a 1D wire, and  remarked that hybridyzation of the photon with the exciton does not necessarily result in a reduction of the coupling constant by the Hopfield factor, as it is commonly assumed in the mean-field framework;
on the contrary, when the spatial correlations are taken into account, the light mass of the photon hinders the minimization  of the repulsion between the polaritons.

\begin{figure*}[t]
    \centering
\includegraphics[width=0.92\linewidth]{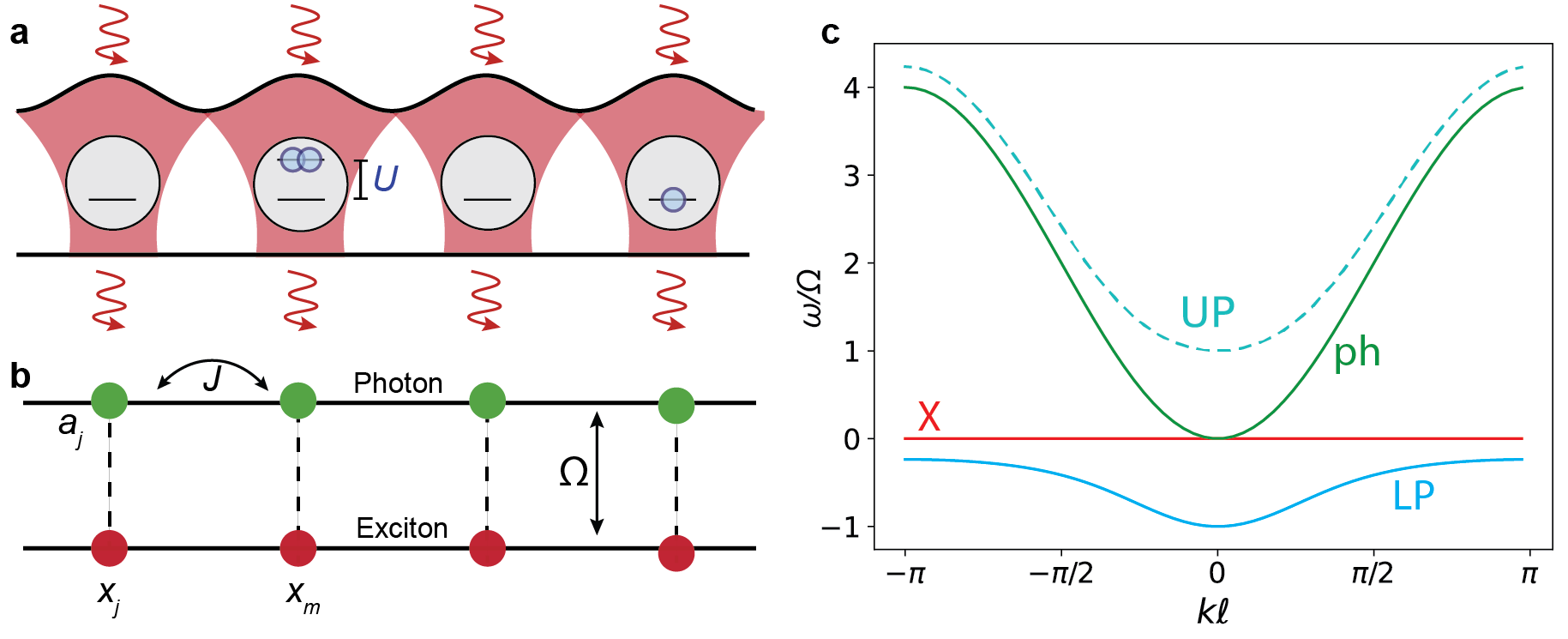}
    \caption{1D lattice of strongly interacting exciton-polaritons. (a) An array of 0D cavity sites, each coupled to a confined excitonic dot. The cavity sites can be individually or collectively pumped. (b) The corresponding ladder model, where the vertical rungs represent the photon-exciton coupling in the system with a coupling strength $\Omega$. Cavity photons hop between sites with a rate $J$, that depends on the cavity design and inter-site distance. The excitonic anharmonicity is given by the on-site interaction strength $U$, which can be enhanced substantially by tighter exciton confinement. (c) The dispersion relations of cavity photons (green), exciton (red) and exciton-polaritons (blue for the lower polariton branch, cyan dashed for the upper one). 
    }
    \label{fig:schematic-figure}
\end{figure*}

In this work we show that, with available 
experimental setups, it is possible to generate a 1D quantum gas of exciton-polaritons displaying the signatures of fermionization in the density-density correlation function and in the dynamic stucture factor, where the Lieb-Liniger hole branch takes a finite weight.
Our proposal consists of an array of excitonic quantum dots coupled to a 1D photonic waveguide or array.
In Sec.~\ref{sec:model} we introduce the underlying microscopic model and discuss the analytical limits of our system.
The roles of the Rabi splitting, photonic bandwidth and excitonic anharmonicity are analyzed in detail, highlighting a crossover between the Tonks-Girardeau and Tavis-Cummings regimes~\cite{tavis_exact_1968}. 
Our numerical results are based on matrix-product state (MPS) calculations and time-evolution techniques, as reported in Sec.~\ref{sec:results}.
In Sec.~\ref{sec:outlook} we summarize our findings and provide an outlook for future investigations.

\section{Setup and Theoretical Model}
\label{sec:model}

In this work we study the ground state and collective excitations of a one-dimensional polariton system, which is schematically depicted in Fig.~\ref{fig:schematic-figure}.
Polaritons possess a matter (exciton) and a radiation (photon) component~\cite{carusotto_quantum_2013}.
In typical experimental realizations, the exciton dispersion is negligible over wavelengths belonging to the light cone of an optical photon.
For this reason we will consider a flat exciton mode in what follows, resonant with the bottom of the 1D photon dispersion. 
The physical system that we envision consists of a 1D array of cavity sites with an active nonlinear material that exhibits a strongly anharmonic spectrum.
We consider an excitonic quantum dot defined by electric fields, where the confinement length is tunable~\cite{Hu2024, Thureja2022,thureja2024electrically}.
When the light-matter coupling strength, represented by the vacuum-Rabi splitting, $\Omega$, is stronger than the overall cavity linewidth ($\Gamma$), we expect 0D exciton-polariton modes in each site (Fig.\,\ref{fig:schematic-figure}\,\textbf{b}).
Interactions between optical photons in vacuum are extremely small, while (polarized) excitons in a same quantum dot repel each other in a sizable way.
Therefore we model local exciton interactions via a Bose-Hubbard term of strength $U$.
For a small enough exciton confinement length in the quantum dot, this interaction-energy shift can exceed the linewidth, leading to polariton blockade.
The 0D cavities are tunnel-coupled to each other, as shown in Fig.\,\ref{fig:schematic-figure}\,\textbf{a}, with a rate $J$, and the dispersion of polaritons in the 1D cavity lattice is shown in Fig.\,\ref{fig:schematic-figure}\,\textbf{c}.

To perform our numerical studies we discretize space and consider a lattice model of length $L$.
As depicted in Fig.~\ref{fig:schematic-figure}\textbf{b}, such a lattice can be thought of as a $L \times 2$ ladder, with the top and bottom rails corresponding to the the photon and exciton degrees of freedom, respectively.
The creation operator for a photon at site $j$ is denoted $a_j^\dagger$, and similarly  $x_j^\dagger$ for the exciton.
Altogether, we consider the Hamiltonian
\begin{multline}
      H = \sum_j \omega_X  x_j^\dagger x_j
        - J \left ( a_j^\dagger a_{j+1} + a^\dagger_{j+1} a_j  
    - 2a_j^\dagger a_j 
    \right) +
    \\
    -
        \Omega \left ( x_j^\dagger a_j + a_j^\dagger x_j \right ) + \frac{U}{2} x_j^\dagger x_j^\dagger x_j x_j
        \label{eq:Hladder}
\end{multline}
where $\omega_X$ denotes the exciton detuning, $J$ is the photon hopping strength, $\Omega$ is the Rabi coupling and $U$ denotes the on-site repulsion between excitons.
In the following, we set the detuning to zero, \textit{i.e.}, $\omega_X=0$.
The bare photon dispersion is given by $\omega_{\rm ph}(k) = 2J(1-\cos(k))$.
In our calculations we fix the lattice constant to one, but below we discuss the role of the physical distance $\ell$ between the quantum dots.
A similar setting was considered in Ref.~\cite{knap_spectral_2010}, which focused on densities close to the Mott transition, \textit{i.e.}, with around one particle per rung.
On the contrary, in the following we will be mainly interested in the dilute regime of small fillings, which displays different physics.

Note that we neglect the driven-dissipative nature of polaritons and deal with a closed system, with a fixed number $N$ of bosonic excitations.
Provided one can engineer a photonic wire with a sufficiently high quality factor, such an approximation may be appropriate for excitons confined to quantum dots, since their oscillator strength is strongly reduced by the dimensional confinement, so their radiative decay rate is very small~\cite{Hamaguchi2023}.
Under such conditions, polaritons may have a sufficiently long lifetime to reach equilibrium, similarly to experiments in ultracold-atom~\cite{anderson_observation_1995,davis_bose-einstein_1995} and photon~\cite{klaers_boseeinstein_2010} BECs.
In our theoretical framework, the non-equilibrium nature of the system  will still enter in the spectral linewidths of the collective excitations displayed below.

We use the following parameters to model our system, that are realistic for excitonic quantum dots (\cite{Hu2024,thureja2024electrically}) in TMD heterostructures coupled to 1D cavity arrays. The tunnel coupling strength between cavity sites is assumed to be in the range of $J \approx (1-5)\,$meV for a $\sim 1 \mu$m distance between optical cavities.
For the on-site interaction strength between polaritons, we estimate values in the range $U \approx (0.1-1)\,$meV.
This is obtained from the typical exciton-exciton interaction strength in TMD monolayers of about $0.1\,\mu\mathrm{eV}\mu \mathrm{m}^2$, which leads to insignificant shifts for delocalized 2D excitons.
However, in electrically defined quantum dots shown in Ref.\,\cite{Hu2024,thureja2024electrically}, the excitonic area can be reduced down to $\sim 100\,\mathrm{nm}^2$.
This in turn dramatically enhances the interaction shift U, potentially more than the typical polariton linewidth $ \Gamma \sim 100\,\mu\mathrm{eV}$.
The reduction of excitonic area also comes at a cost, i.e. the reduction of the light-matter coupling.
However, the large oscillator strength of excitons in TMD monolayers is of great advantage in this case.
The vacuum Rabi splitting for TMD monolayers coupled to microcavities has been shown to be up to $\sim 20\,$meV \cite{tan2020}.
For electrically defined quantum dots with reduced area, we estimate $\Omega \sim (2-5)$\,meV, which also exceeds the linewidth.
Therefore, with these parameters, the scenario of strongly interacting polaritons with on-site blockade can be realized. 

In the next paragraphs, we build some physical intuition for the nature of the ground state of the system.
To this aim, we study the two limiting regimes of small and large $J/\Omega$, displaying the physics of polariton fermionization and of the Tavis-Cummings model, respectively.

\subsection{Tonks-Girardeau limit}

In the limit of small $J/\Omega$ or small density, the kinetic energy of the system is negligible.
As a consequence, the ground state optimizes the light-matter interaction and consists of well-defined (lower-branch) polaritons.
Because of the zero exciton-photon detuning, such polaritons will be half matter-like and half radiation-like.
In other words, the Hopfield coefficient will be $0.5$.
We can thus expect that the effective hopping of a polariton is approximatively given by $J_{\rm pol} = J/2$.
On the other hand, polaritons inherit interactions from their excitonic component.
As recently demonstrated in Ref.~\cite{christensen_microscopic_2022}, the correct way to extract the effective polariton-polariton interaction constant $U_{\rm pol}$ is to adopt a Born-Oppenheimer approach: one neglects the kinetic energy and computes the ground-state energy $E_{BO}$ of 
\begin{equation}
H_{BO} =
    \begin{pmatrix}
    0 & -\sqrt{2} \Omega & 0 \\
    -\sqrt{2} \Omega & 0 & -\sqrt{2} \Omega \\
    0 & -\sqrt{2} \Omega & U
    \end{pmatrix},
\end{equation}
corresponding to the  matrix elements of the Rabi coupling plus exciton-exciton interactions in the two-boson sector. 
Then, one defines $U_{\rm pol} = E_{BO} - 2 \Omega$, which is the two-body local energy difference between the interacting and non-interacting cases.
In particular, when $U/\Omega \to 0$ one recovers $U_{\rm pol}=U/4$, which is the expected result when taking the Hopfield coefficient to the fourth power, while for $U/\Omega \to \infty$ one has $U_{\rm pol} = (2-\sqrt{2})\Omega$.
In summary, the effective lower-polariton Hamiltonian reads
\begin{multline}
    H_{\rm pol}
    =
    -J_{\rm pol}
    \sum_j (b^\dagger_j b_{j+1} + b^\dagger_j b_{j-1})
        +
    \\
    +
    (2J_{pol}-\Omega) \sum_j b^\dagger_j b_j
    +
    \frac{U_{\rm pol}}{2} \sum_j b^\dagger_j b^\dagger_j b_j b_j,
    \label{eq:Hpol}
\end{multline}
where 
$b_j^\dagger = \frac{a_j^\dagger + x_j^\dagger}{\sqrt{2}}$ describes the creation of a lower branch polariton.

\begin{figure*}[t]
    \centering
    \includegraphics[width=0.32\linewidth]{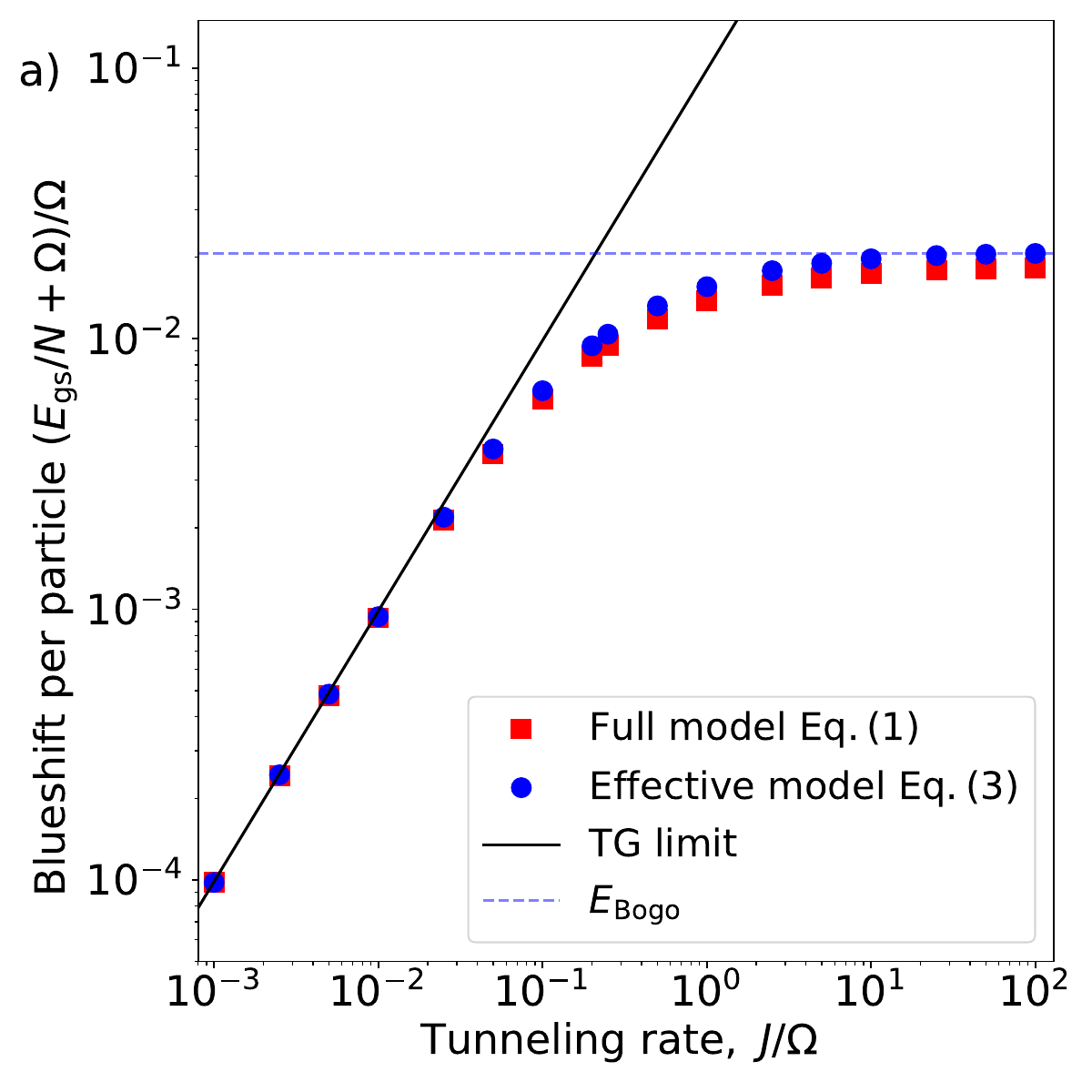}
    \includegraphics[width=0.32\linewidth]{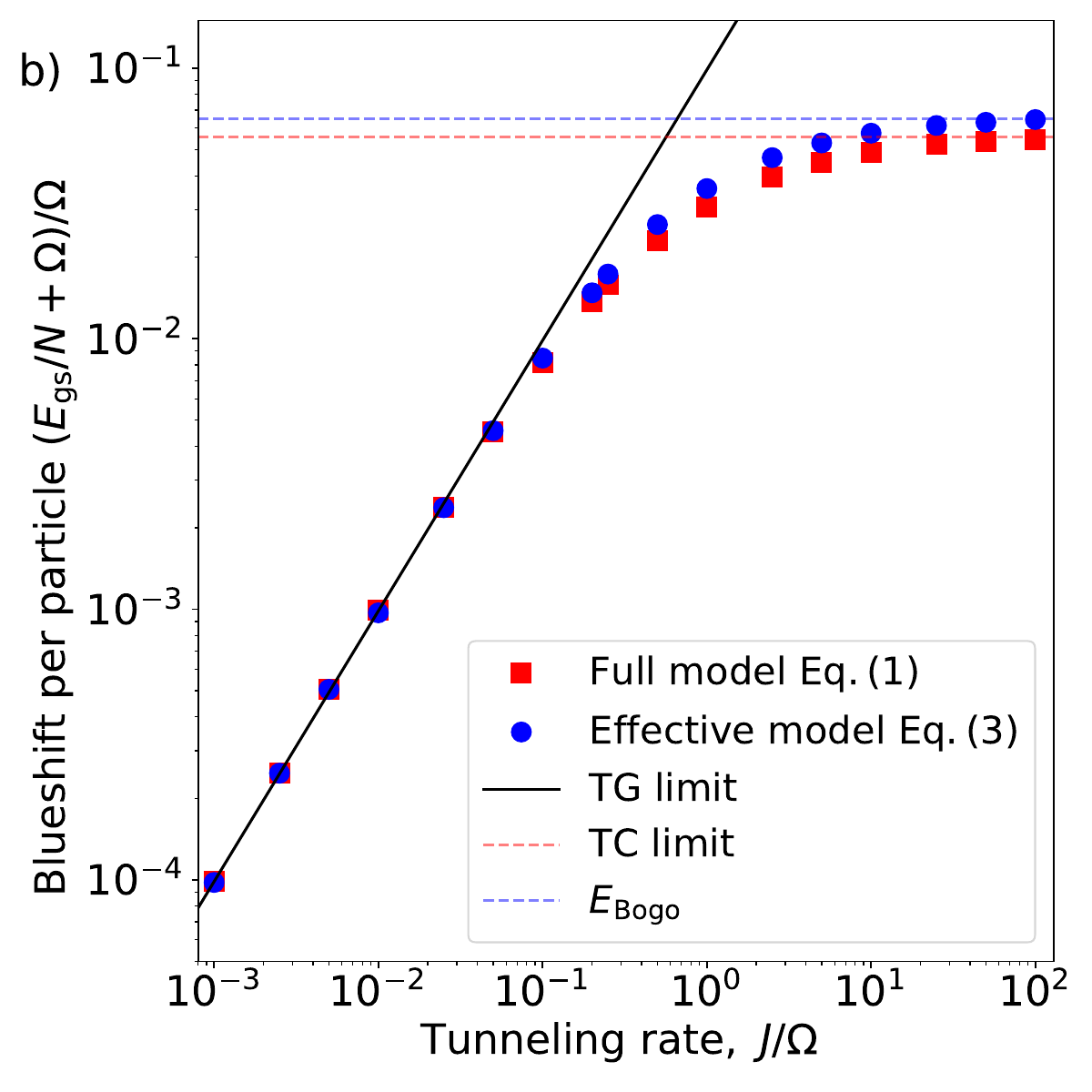}
    \includegraphics[width=0.32\linewidth]{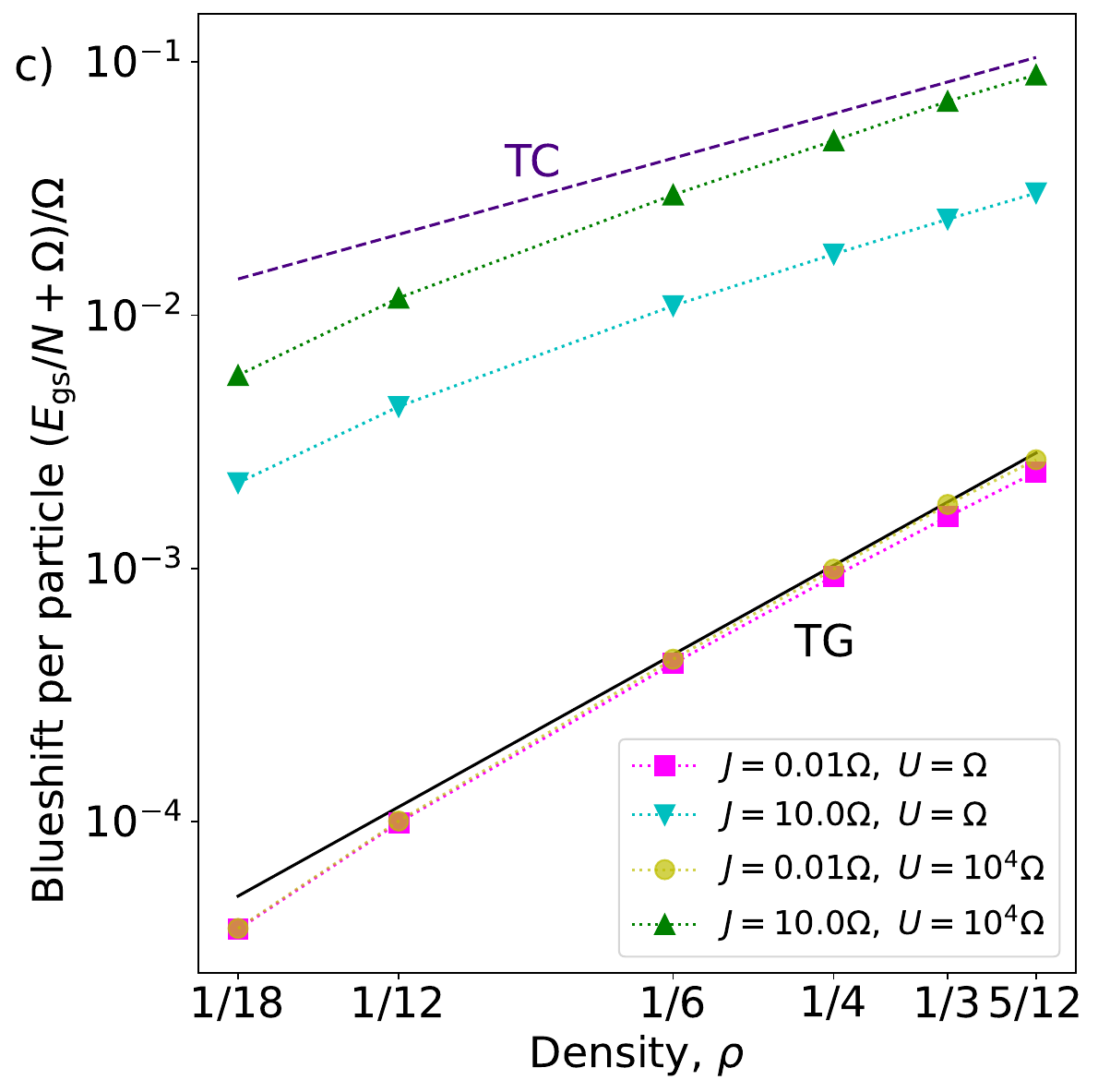}
    \caption{Blueshift per particle of the ground-state energy with respect to the non-interacting case, for a periodic system of size $L = 36$.
   Fixing $N=9$ excitations, the behavior as function of $J/\Omega$ has been analysed for (a) realistic $U=\Omega$  and (b) very large $U=10^4\Omega$ exciton-exciton interactions.
   At small $J/\Omega$ the onset of fermionization is indicated by the black solid line.
   In panel (b) for very large $J/\Omega$ one achieves the Tavis-Cummings limit (red dashes).
   The full photon-exciton model (red squares) is also compared with an effective lower polariton model (blue circles), displaying perfect agreement at small $J/\Omega$.
   In panel (c) we instead vary $N$, hence the density $\rho$ of excitations.}
    \label{fig:blueshift}
\end{figure*}

For $J \ll \Omega,U$, which entails $J_{\rm pol} \ll U_{\rm pol}$, the physics is then well captured by the lattice version of the Tonks-Girardeau (TG) model \cite{tonks_complete_1936,girardeau_relationship_2004}.
In this scenario, polaritons behave like impenetrable bosons and they can be mapped onto fermions by the usual Jordan-Wigner transformation.
As a result, any eigenstate of the bosonic system is in one-to-one correspondence with a 1D non-interacting Fermi gas, with the same energy and total momentum.
(Other quantities, like the momentum distribution, require instead a highly non-trivial mapping.)
The ground state, in particular, simply corresponds to a Fermi sea filled up to the Fermi momentum $k_F = \pi N/L$.
More precisely, for $N=2M+1$ particles  and with periodic boundary conditions (PBC), the TG energy reads 
\begin{equation}
    E_{TG} = -N\Omega 
    +
    J_{\rm pol} \sum_{m=-M}^M
    \left[ 
    2 - 2\cos\left(\frac{2\pi}{L}m \right)
    \right],
    \label{eq:E_TG_discrete}
\end{equation}
where $-N \Omega$ is the exact energy at $J=0$.
For an even number of particles, the ground state is instead degenerate at the TG level.

The Fermi momentum $k_F = \pi N/L$ plays a fundamental role in determining the collective excitations and the extension of the Lieb-Liniger branch II~\cite{lieb_exact_1963}, which will be analyzed in Sec.~\ref{ssec:tavis-cummings} and basically reflects the fact that the scattering of a particle between the two Fermi surface points costs zero energy and involves a $2k_F$ momentum exchange. 

In a continuum wire, or in the limit of small densities, the TG scaling holds:
\begin{equation}
    E_{TG}/N = -\Omega 
    +  \frac{k_F^2}{6m_{\rm pol}} 
    \label{eq:E_TG_continuum}
\end{equation}
where now the Fermi momentum  reads
$k_F = \pi N/(\ell L)$ and the polariton mass $m_{\rm pol}^{-1} = {2 \ell^2 J_{\rm pol}}$.

\subsection{Tavis-Cummings regime \label{ssec:tavis-cummings}}

At large $J$ and with PBC, the only state accessible for the photons is the $k=0$ mode.
In the limit $U/\Omega \to \infty$, one can approximate the the excitonic Hilbert space at each site as a two-level system, to recover the Tavis-Cummings physics of collective light-matter interaction~\cite{tavis_exact_1968}, given by the Hamiltonian
\begin{equation}
    H_{TC} =
    -\frac{\Omega}{\sqrt{L}} a_0^\dagger \sum_j^L 
    x_j,
\end{equation}
to be solved in the subspace with $N$ excitations.
This Hamiltonian can be block-diagonalized using the Dicke representation of angular momenta~\cite{mandel_optical_1995} 
\begin{equation}
    S_+ = \sum_j^L 
    x^\dagger_j, \ \ \ 
    S_z = \sum_j^L 
    \left( x^\dagger_j x_j - \frac{1}{2} \right),
\end{equation}
obeying the algebra $
    [S_+,S_-] = 2 S_z, \ \ \ 
    [S_z,S_\pm] = \pm S_\pm
$.
The ground state lives in the  representation of maximal permutation symmetry or, in other words, of maximal Dicke angular momentum $S=L/2$, where the energy can be easily evaluated numerically.
At small filling $\rho = N/L \ll 1$, it holds that
\begin{equation}
    E_{TC}/N \simeq  - \Omega + \frac{\Omega}{4} \rho 
    + O\left( \rho^2 \right).
    \label{eq:E_TC}
\end{equation}
Notice that this is independent of $J$ and entails that the crossover between the TG and TC regimes occurs for $J \sim \frac{3\Omega}{2\pi^3\rho^2}$.
In other words, having a larger density favors the TC phase and vice versa.

At large $J$ but small $U/\Omega$, we instead expect that the Bogoliubov approximation for the polariton Bose gas  should become accurate, yielding a ground-state energy of 
$E_{\rm Bogo} \simeq -\Omega N + \frac{U_{\rm pol}}{2} \frac{N(N-1)}{L}$.
In this regime, one has no fermionization and the low-energy excitations will be dominated by the usual bosonic Bogoliubov dispersion~\cite{pitaevskii_bose-einstein_2016}, with the acoustic Goldstone mode at small momenta and no Lieb-Liniger branch II.

\begin{figure*}[t]
    \centering
    \includegraphics[width=0.48\linewidth]{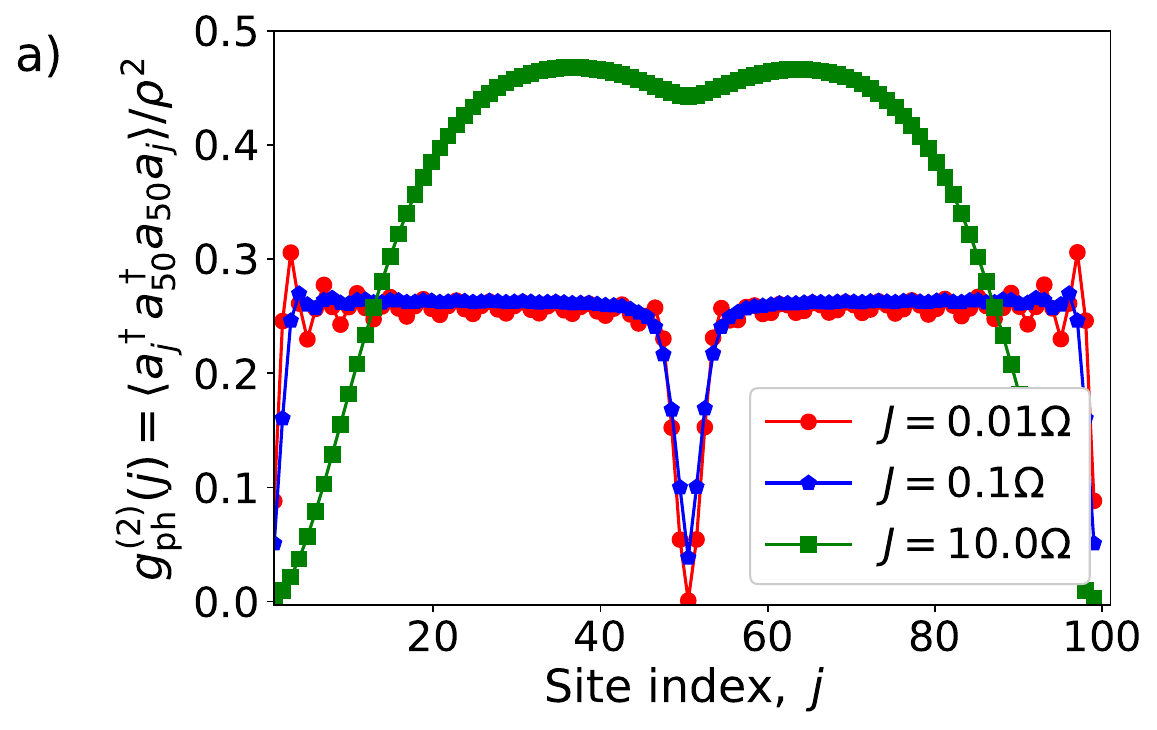}
    \includegraphics[width=0.48\linewidth]{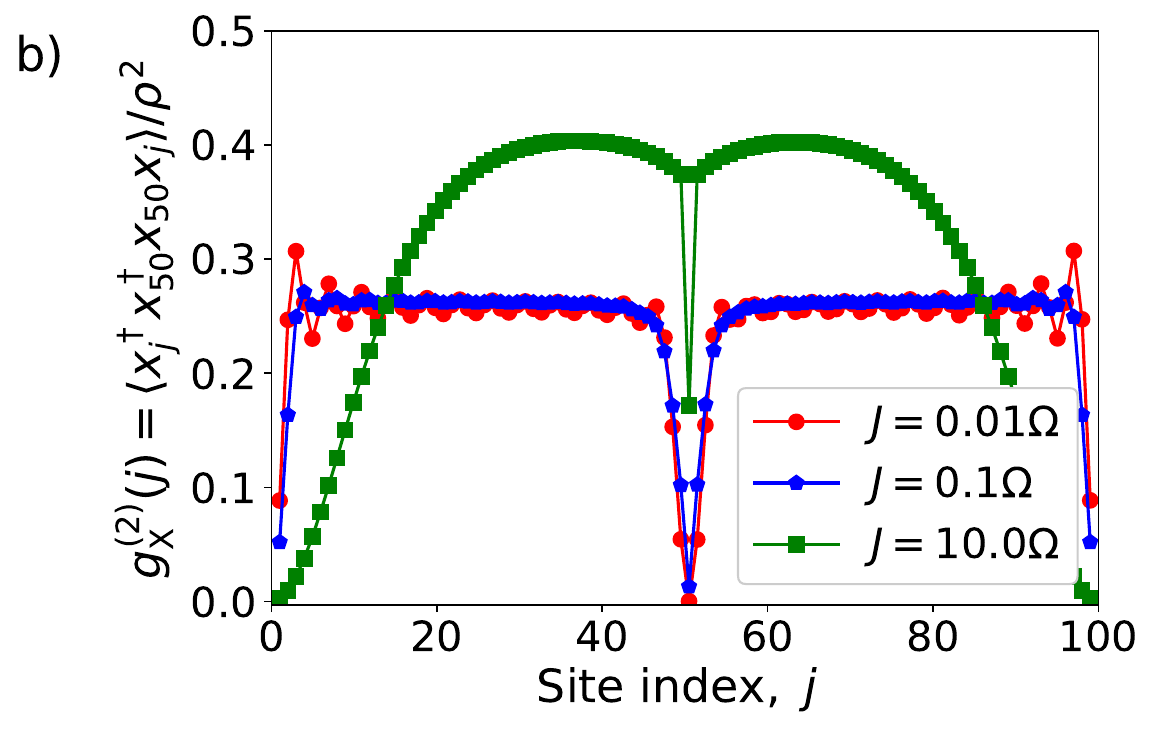}
    \caption{(a) Photonic and (b) excitonic $g^{(2)}$
    functions, evaluated with respect to the central site of an OBC ladder. Here we used a realistic $U=\Omega$ interaction strength, $N=25$ excitations in $L=100$ sites (one-quarter filling)
    and three different hopping couplings $J/\Omega$.
    At $J=0.01 \Omega$ (red circles), Friedel oscillations are clearly visible and provide a hallmark of fermionization.}
    \label{fig:g2}
\end{figure*}

To conclude these theoretical remarks, let us discuss the case of open boundary conditions (OBC).
For small hopping $J$ and large lattice length $L$, the physics discussed above for PBC is recovered in the bulk of the system, with small deviations at the endpoints (see App.~\ref{app:finite-size}).
For a finite $L$ and very large $J$, one would instead find that the only accessible photonic mode is a sinusoidal wave with finite energy.
The excitations will then tend to avoid the photonic sector and the ground state becomes matter-like (see App.~\ref{app:photonic-fraction}).
In the following we will consider PBC for the computation of the energies, since we want to show its scaling at arbitrary $J$, while for the excitation spectra we will use OBC, because in our MPS-based numerical calculations it is cheaper and much easier to treat (even rather large, $L\gtrsim 100$) OBC systems than (intermediate-scale, $L\gtrsim 50$) PBC systems.
In our OBC calculations, we will therefore only consider relatively small $J$ and large $L$, such that we only deal with bulk physics, and boundary and finite-size effects become negligible.

\section{Numerical results}
\label{sec:results}

As already mentioned, there is a double advantage in setting up the system on a lattice.
The first, physical, reason is that this allows to tune the photon hopping $J$ by the design of the photonic structure.
The second, purely theoretical, advantage is that we can profit from the power of MPS-based methods such as DMRG to numerically study the ladder model of Eq.~(\ref{eq:Hladder}).

\subsection{Ground-state energy}

In Fig.~\ref{fig:blueshift} we investigate the ground state energy by defining the blueshift per particle, $E_{GS}/N+\Omega$, which represents the energy per particle minus the energy of a free polariton, $-\Omega$, thus quantifying the effect that interactions have on the total energy.

In Figs.~\ref{fig:blueshift}a) and \ref{fig:blueshift}b) we fix the filling factor to one quarter and consider $N=9$ excitations in a lattice of length $L=36$.
In Figs.~\ref{fig:blueshift}a) we numerically vary $J/\Omega$ for the realistic interaction strength $U=\Omega$.
The DMRG results for the ladder model are reported as red squares.
As anticipated, the polaritons are effectively impenetrable at small $J/\Omega$ and one recovers the Fermi sea energy of the TG gas, Eq.~\eqref{eq:E_TG_discrete}, indicated by a black solid line.
In this regime, the effective polariton model from Eq.~(\ref{eq:Hpol}) reproduces perfectly the full result, as indicated by the blue circles, obtained by running DMRG on the effective model.
At large $J$ there is instead some discrepancy, since with $U \sim J \sim \Omega$ one cannot exclude some ionization amplitude for the polariton.
The effective model is well captured by the Bogoliubov result, $E_{\rm Bogo} \simeq -\Omega N + \frac{U_{\rm pol}}{2} \frac{N(N-1)}{L}$ (see the blue dashed line) suggesting the onset of quasi-condensation.
Here, $U_{\rm pol} \simeq 0.186 \ \Omega$ is used, as obtained from the Born-Oppenheimer approximation. 

Fig.~\ref{fig:blueshift}b) reports the same calculation but for the sigificantly larger interaction strength $U=10^4 \Omega$, which effectively means that excitons are hard-core bosons.
At small $J$, the Fermi sea energy is recovered and the effective model, where now we set 
$U_{\rm pol}
= (2-\sqrt{2})\Omega$, works perfectly.
At large $J$, instead, the blueshift matches the Tavis-Cummings prediction of Eq.~(\ref{eq:E_TC}), see red dashes,
and thus reveals the physics of collective light-matter interaction of a single photonic mode with a collection of two-level emitters.
The effective model energy, instead,
for large $J/\Omega$  is once again grasped by the Bogoliubov approximation.

In Fig.~\ref{fig:blueshift}c) we show the behavior of the blueshift per particle at varying densities $\rho = N/L$ and for a few combinations of $U$ and $J$.
In particular, it is clear that when $J \ll U,\Omega$ the TG physics is recovered and is independent of $U$, since it can be mapped to the non-interacting Fermi sea.
This is evident by looking at the perfect overlap of the yellow circles and fuchsia squares (at least at small densities).
The black solid line corresponds to $E_{TG}/N$ from Eq.~\eqref{eq:E_TG_continuum} and scales with density as $ \propto \rho^2$.
The small discrepancies are finite-particle and finite-density deviations from the continuum line.
When $J/\Omega$ is large, the dominant scaling of the single-particle blueshift is $ \propto \rho$, but the results are sensitive to the value of $U$.
In particular, only for very large $U$ and $J$ the TC prediction is recovered (see the purple dashed line corresponding to Eq.~\eqref{eq:E_TC}).

\begin{figure*}[t]
    \centering
\includegraphics[width=0.98\linewidth]{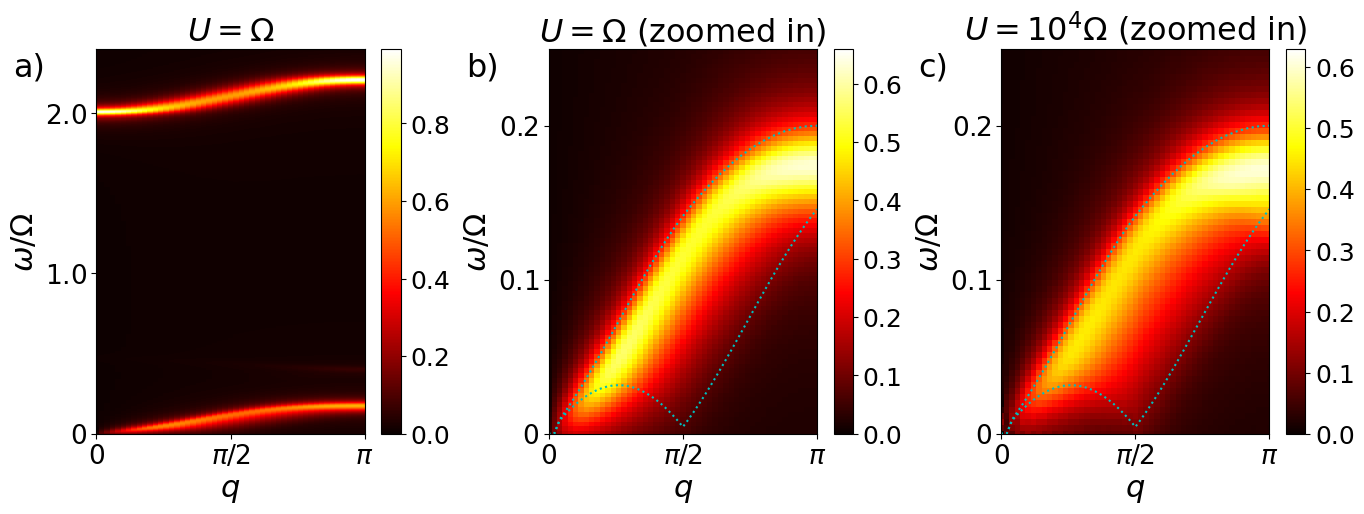}
    \caption{
    Dynamic structure factor $S(q,\omega)$
    for a system of $N=25$ excitations in a $L=100$ chain,
    with photon hopping $J=0.1\Omega$.
    Panel (a) reports $S(q,\omega)$
    for a realistic $U=\Omega$
    interaction strength
    and features three main contributions: the high energy one is due to Rabi splitting, the weak intermediate line to doublons, the low energy one consists of phonons and of the Lieb-Liniger II excitations.
    A zoom of this latter part is provided in panel (b), where the area containing the fermionic excitations is limited by the cyan dashed line. 
    Panel (c) reports the large $U=10^4\Omega$ case, where the Lieb-Liniger branch II is more pronounced.
    For the numerical time evolution, we use a step size of $\Omega \cdot \delta t = 0.1$ and a total propagation time $\Omega T=500$.}
    \label{fig:Swq}
\end{figure*}

These calculations have been performed with PBC and using $L=36$ sites; we refer to App.~\ref{app:finite-size} for a finite-size analysis, confirming that finite-size corrections are small and well under control.

\subsection{Ground-state correlations}

Another useful observable is given by the second-order correlation functions, strictly related to the density-density correlators.
These tell us about the probability of measuring a second particle at site $j$ given a first particle at site $k$.
Since in our system we have two species of bosons we can consider both the photonic correlator
\begin{equation}\label{eq:photonic-correlator}
 g_{\rm ph}^{(2)}(j,k) = \frac{\langle
a^\dagger_{j} a^\dagger_k a_k  a_j
 \rangle}{\rho^2},
\end{equation}
with $j,k=1,2,...,L$,
and the excitonic one
\begin{equation}
 g_{\rm X}^{(2)}(j,k) = \frac{\langle
x^\dagger_{j} x^\dagger_k x_k  x_j
 \rangle}{\rho^2}.
\end{equation}
Notice that we chose to normalize with respect to the average total density $\rho$  (in contrast with the local  species-resolved one).
Also, in view of the OBC calculations of the next paragraphs, here we report OBC results with $L=101$
and thus we keep the two spatial indices $j,k$ in the $g^{(2)}$ function
(as opposed to a PBC calculation, where only the distance $|j-k|$ would matter).
In practice, we computed the density-density correlations using DMRG and obtained the $g^{(2)}$ functions by performing the commutator. 

In Fig. \ref{fig:g2}a) we report the photonic correlator $g_{\rm ph}^{(2)}(j)$ evaluated in the $N=25, U=\Omega$ ground state and having fixed $k=j_0=50$ at the central site of the chain.
For small $J/\Omega=0.01$ (red circles) the effect of fermionization is evident from the strong Friedel oscillations, originating from both the reference sites and the edge of the system.
For larger $J/\Omega=0.1$ (blue pentagons) the Friedel oscillations have the same period but are very weak.
Notice that, in spite of the relatively small system size, the bulk value of the $g^{(2)}$ function approaches $1/4$, coming from the fact that the photonic content of the GS is around 50\%.
For larger $J/\Omega=10.0$ (green squares)
the Friedel oscillations are washed out
and the shape of 
$g^{(2)}$
is dominated by the sinusoidal lowest energy photonic mode.
In this regime, the system size as well as the choice of boundary conditions play a major role, and we are far from being in the thermodynamic limit.
Also, notice that the photonic fraction is no longer close to 50\%, see App.~\ref{app:finite-size}.

Furthermore, the correlators show a clear dip at $j=k$.
In the TG regime the polaritons are impenetrable and the $g^{(2)}$ function is vanishing.
For large $J$ and with OBC, the light and matter degrees of freedom decouple, as also shown by the Von Neumann entropy in App.~\ref{app:von-neumann}.
This means that two or more photons have a finite probability to be found at the same lattice site.

Fig. \ref{fig:g2}b) displays the excitonic correlation function, $g_{\rm X}^{(2)}(j)$.
At small $J/\Omega$ there is no big difference with respect to the photonic correlator, confirming that the polaritons are well-bound particles.
Instead, at large $J/\Omega$ the dip at $j=k$ survives, even though the large kinetic energy in the system makes it possible to find two excitons at the same spot.
Notice that for $U \to \infty$ the excitonic dip always reaches zero.

\subsection{Dynamic structure factor}

\begin{figure*}[t]
    \centering
\includegraphics[width=0.98\linewidth]{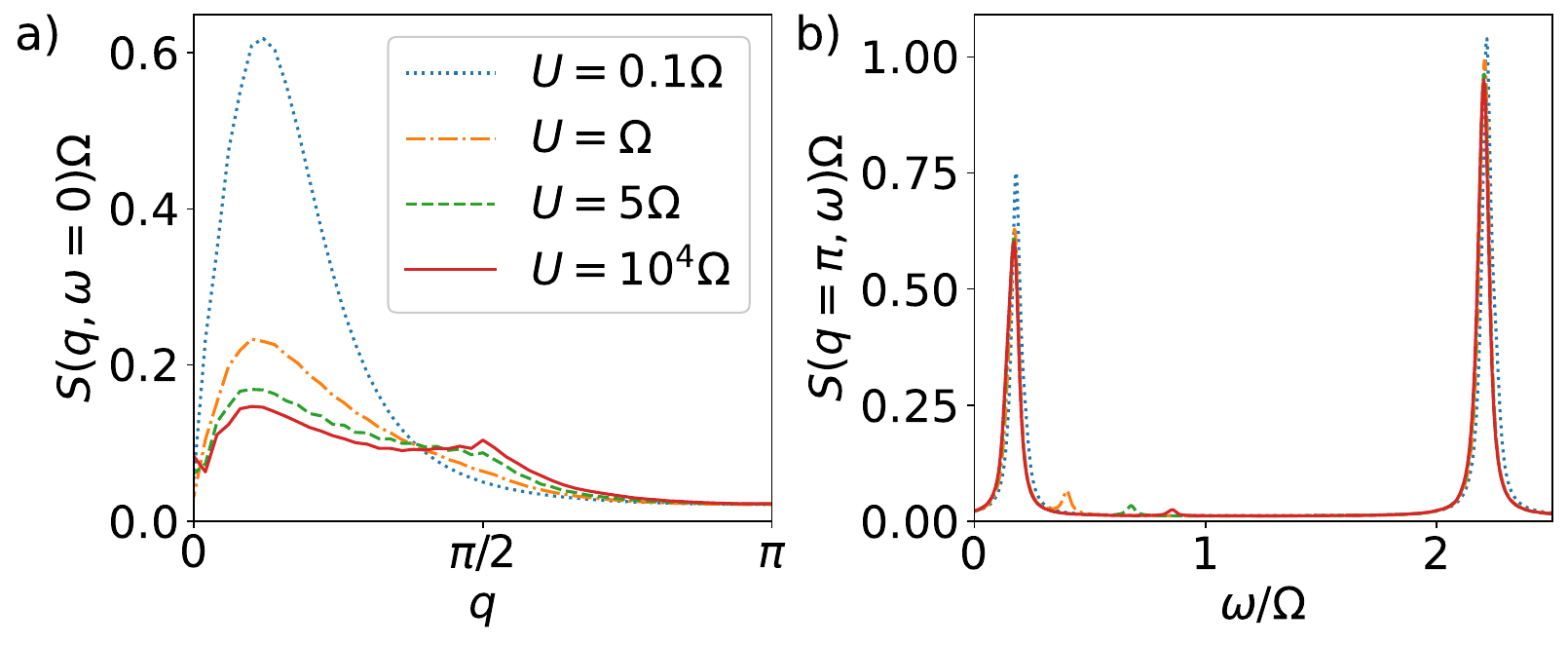}
    \caption{Slices of the dynamic structure factor $S(q,\omega)$
    are reported for different $U$'s, and other parameters as in Fig.~\ref{fig:Swq}.
    In panel (a) we slice along $\omega=0$ and show that interactions give rise to a small peak around $q=2k_F \sim \pi/2$
    \label{fig:S_cut} and determine a loss of oscillator strength at small $q$.
    In panel (b) we cut at $q=\pi$ to highlight the presence of a set of weak doublon peaks shifting with $U_{\rm pol}$, and disappearing at negligible $U$.
    }
    \label{fig:Swq-cuts}
\end{figure*}

The momentum and frequency-dependent response of a system's density to external perturbations is encoded in the dynamic structure factor. 
It provides insights into the collective neutral modes of the system; a paradigmatic, historically crucial, example is provided by the acousto-rotonic dispersion  of phonons in superfluid helium, as probed by inelastic neutron scattering~\cite{donnelly_specific_1981}.
We compute the dynamic structure factor of channel $\alpha \in \{ {\rm ph,X} \}$ as
\begin{equation}\label{eq:sqw}
    S_{\alpha\alpha}(q,\omega)
    =
    {\rm Re}
    \sum_j
    \int_0^{+\infty} \frac{dt}{\pi} \ e^{i\omega t-iqj} \langle \delta n_{\alpha}(j,t) \delta n_{\alpha}^\dagger(j_0,0) \rangle
\end{equation}
where we introduced the density-fluctuation operator 
$\delta n_\alpha
=
n_\alpha - \langle
n_\alpha
\rangle
$.
To get a broadening of the lines $\Gamma$, we just multiply by $e^{-\frac{\Gamma}{2} t}$ before taking the Fourier transform, and
we use $\Gamma=0.04  \Omega$, as an estimate of the polariton radiative rate.
In view of the experimental detection, we will focus on 
the $\alpha={\rm ph}$ sector and drop the subscript $\alpha\alpha$
in the following.

We use density matrix renormalization group (DMRG) calculations \cite{white_density_1992,fishman_itensor_2022} and the time-dependent variational principle \cite{paeckel_time-evolution_2019} to obtain the time evolution of $n_{j_0}|GS\rangle$ under the full Hamiltonian in Eq.~\eqref{eq:Hladder}.
According to Eq.~\eqref{eq:sqw} this yields access to the dynamic structure factor upon Fourier transformations.
The results are reported in Fig.~\ref{fig:Swq}, for an OBC chain of length $L=100$ and $J=0.1\Omega$.
At $q=0$ we find that the relevant spectral information in the structure factor are superimposed by numerical artefacts that stem from a phase shift in the time-evolved state.
While this error may be reduced to some extent by improving the time-evolution method \cite{paeckel_time-evolution_2019}, it is not a major concern for us and we instead set the corresponding column of $S(q,w)$ to zero.
In panel (a), one can distinguish two main branches, and a fainter one in between.
The high-energy branch corresponds to excitations from the lower polariton branch into an upper polariton branch, and is located at energies of order $2\Omega$.
As it will be discussed below,  the intermediate and weakest branch has its position determined by $U_{\rm pol}$ and originates from doublon excitations. 

The low-energy band is instead zoomed in in panel (b) and bears the hallmarks of fermionization.
The mapping between impenetrable 1D bosons and a non-interacting Fermi gas entails 
a one-to-one correspondence between all the eigenstates, where energy and total momentum are preserved in the mapping.
There are then two possible kinds of excitations with very small energy.
The first one is the excitation of a fermion close to the Fermi surface with a small momentum transfer.
The second one is the excitation of a fermion by a momentum transfer of order $2k_F$.
In the bosonic system, this results in the presence of two branches in the dynamical structure factor:
the usual acoustic phonon dispersion at small momenta, and the so-called Lieb-Liniger branch II at momenta around $2k_F$.
The two branches are displayed in Fig.~\ref{fig:Swq}b) by the cyan dashed lines.
The area between the two lines comprises all the $(q,\omega)$ points for which the Lindhard function of the free Fermi gas is non-zero.
In panel (c), we also report the case of very large $U=10^4 \Omega$. 

While in the color plots of Fig.~\ref{fig:Swq} the presence of the Lieb-Liniger branch II is barely visible, the fingerprints of fermionization are more evident when looking into slices of the dynamic structure factor at different values of $U$.
In Fig.~\ref{fig:S_cut}a), we report the cut of $S(q,\omega)$ at zero frequency, as a function of momentum and for different $U$'s.
For increasing $U$, the Lieb-Liniger II peak develops around $2k_F\simeq \pi/2$, the filling being one fourth here.
While this phenomenon is very weak for the $U=\Omega$ case (orange dash-dotted line), there is still a large oscillator-strength loss with respect to the non-interacting polariton case. 
Here, one can exactly compute that the low-energy part of $S(q,\omega)$ should be a perfect cosine with 
oscillator strength independent of $q$.
For a numerical comparison, the case $U=0.1\Omega$ is reported as a blue dotted line, showing a qualitative change in the strength and spread of the spectrum.

Another hallmark of interactions is shown in Fig.~\ref{fig:S_cut}b), where we display $S(q,\omega)$ at $q=\pi$.
The small, intermediate peaks are shifted from the low-energy branch by  $\sim U_{\rm pol}$,
suggesting this feature is due to doublon excitations.
Also, the fact that this intermediate branch is more visible at $q \sim \pi$ suggests a lattice origin, and indeed doublons are not well defined in the continuum.

\subsection{Resonant spectroscopy}

\begin{figure}[t]
    \centering
\includegraphics[width=0.99\linewidth]{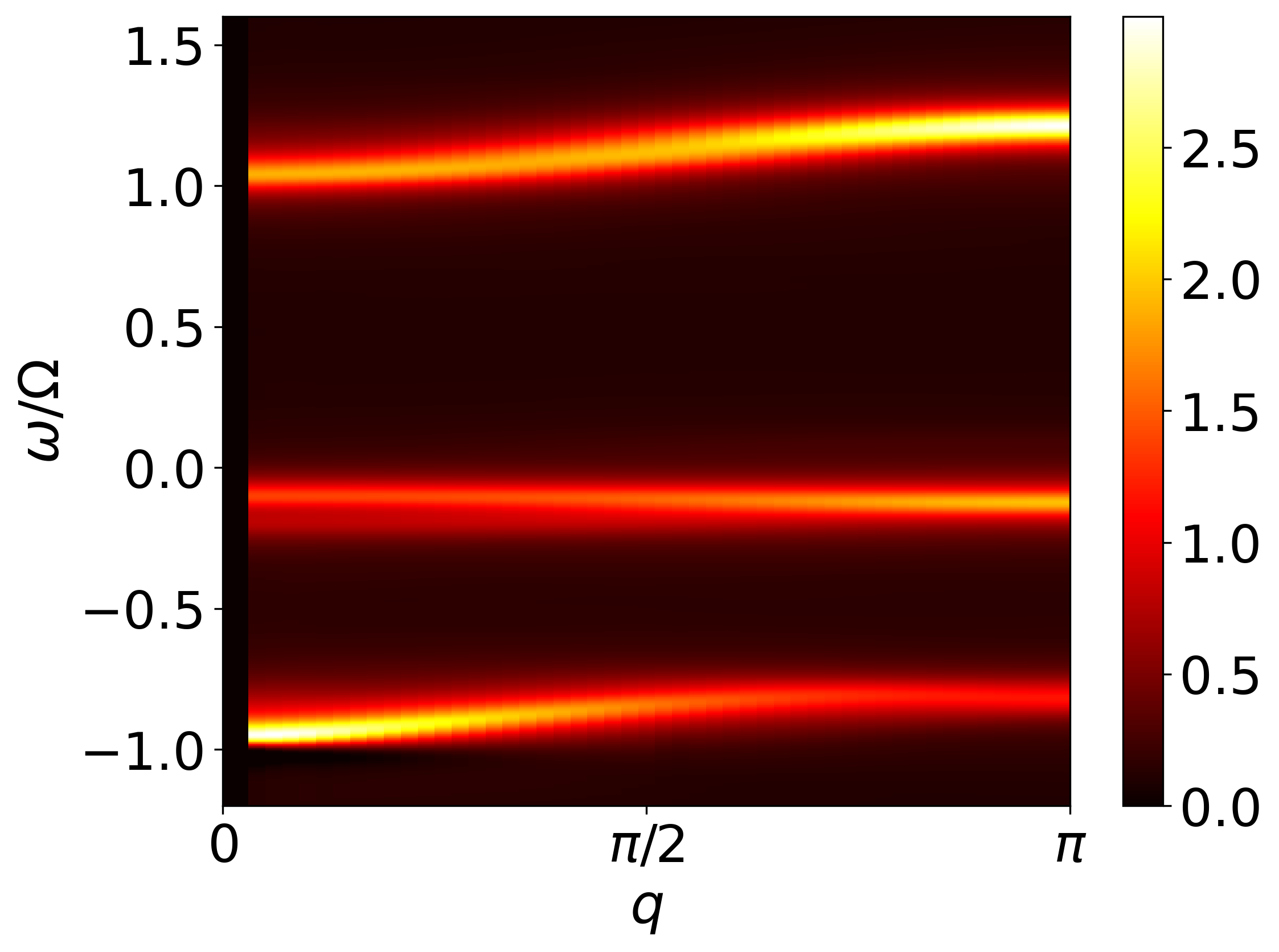}
    \caption{
    Response function $-{\rm Im} \mathcal{\chi}_{\rm res}(q, \omega)$
    for the same parameters as used in Fig.~\ref{fig:Swq}.c,which can be probed by transmission and reflectivity spectroscopy.
    }
    \label{fig:chi_aA}
\end{figure}

In polariton experiments optical spectroscopy enables
 measurements
 like reflectivity, transmittivity, photoluminescence and absorption.
 These possibilities make polaritons a very convenient platform for their easiness to probe.

In particular, in reflectivity and transmittivity measurments the 
cavity is actively perturbed by a probe laser
and
the following response function determines the spectrum~\cite{sieberer_keldysh_2016,caleffi_collective_2023}:
\begin{equation}
    \mathcal{\chi}_{\rm res}(q, \omega) = 
    - i
    \int_0^{+\infty} \frac{dt}{\pi} \
    e^{i\omega t}
    \langle
[a_q(t), a_q^\dagger(0)]
    \rangle,
    \label{eq:chi_aA}
\end{equation}
where $a_q = \frac{1}{\sqrt{N}}\sum_j 
e^{-iqj}
a_j $.
Absorption and photoluminescence experiments correspond instead to the first and second half of the commutator, respectively~\cite{szymaifmmode_acutenelse_nfiska_nonequilibrium_2006}.

The response  function $\mathcal{\chi}_{\rm res}(q, \omega)$
can be once again computed by time evolving the matrix product ground state after addition or removal of a photon.
We plot the resulting
$-{\rm Im} \mathcal{\chi}_{\rm res}(q, \omega)$
in Fig.~\ref{fig:chi_aA}.
One can distinguish three main branches, which we attribute, from lower to higher energy, to the injection of a lower polariton, of a doublon (out of a single polariton), and an upper polariton.
Interestingly, while in principle the single-particle spectral function of the TG gas should also carry the signatures of fermionization~\cite{settino_exact_2021}, in the photon spectral function this is basically invisible (there is only an extremely weak kink around $q \simeq 2k_F$, not visible on this scale).
We conclude that in practice it is easier to infer fermionization from the density-density correlations than by resonant photon spectroscopy.

\section{Summary and outlook}
\label{sec:outlook}

Exciton-polariton gases confined to low dimension offer new experimental routes for studying strongly correlated quantum systems.
Here, based on a microscopic lattice model used to characterize the system, we analyzed the competition between the Rabi splitting, photonic bandwidth and exciton repulsion.
We have 
investigated the 
conditions, in terms of exciton anharmonicity and photonic bandwidth, under which polaritons display fermionized behavior, 
demonstrating  that this regime may be probed by state-of-the-art setups.
On the analytical side, we identified two important limiting cases of the physical parameters,  revealing Tonks-Girardeau and Tavis-Cummings physics, respectively.
Our numerical analysis has shown that it may be favourable to probe fermionization based on density-density correlations rather than by resonance spectroscopy.
With these findings, our work motivates experiments to observe  correlated exciton-polaritons in the near term.

\section*{Acknowledgements}
We are grateful to Iacopo Carusotto for helpful feedback on the manuscript.

J.K. gratefully acknowledges support from Dr. Max Rössler, the Walter Haefner Foundation and the ETH Zürich Foundation. 
I.A. was financially supported by the ERC grant LATIS, the EOS project CHEQS and the FRS-FNRS (Belgium). P.A.M. acknowledges support from Swiss National Science Foundation (SNSF) Starting Grant No. 211448. 

\appendix

\section{Finite-size effects \label{app:finite-size}}

In the main text we report numerical simulations on systems with both periodic and open boundary conditions, the latter configuration more suitable for computing  correlators, as plotted in
Figs. 3--6 at small $J/\Omega$.
Instead, in Fig.~\ref{fig:blueshift},  we report the ground state energy with periodic boundary conditions.
Here we show that, for sufficiently small $J/\Omega$ and sufficiently large $L$, our OBC simulations recover the essential physics occurring the bulk of the system.

\subsection{Ground-state energies}

In Fig.~\ref{fig:blueshift} we have shown DMRG results for the blueshift per particle, $(E_\mathrm{gs}/N+\Omega)/\Omega$, as a function of tunneling rate $J/\Omega$ and for a system with periodic boundary conditions.
Here we instead fix $J/\Omega = 0.1$ and study the blueshift as a function of lattice size $L$.
While doing so, we keep the density, $\rho = N/L$, constant.
As in Fig.~\ref{fig:blueshift}, we choose a filling factor of $1/4$.

\begin{figure}[b!]
    \centering
\includegraphics[width=0.99\linewidth]{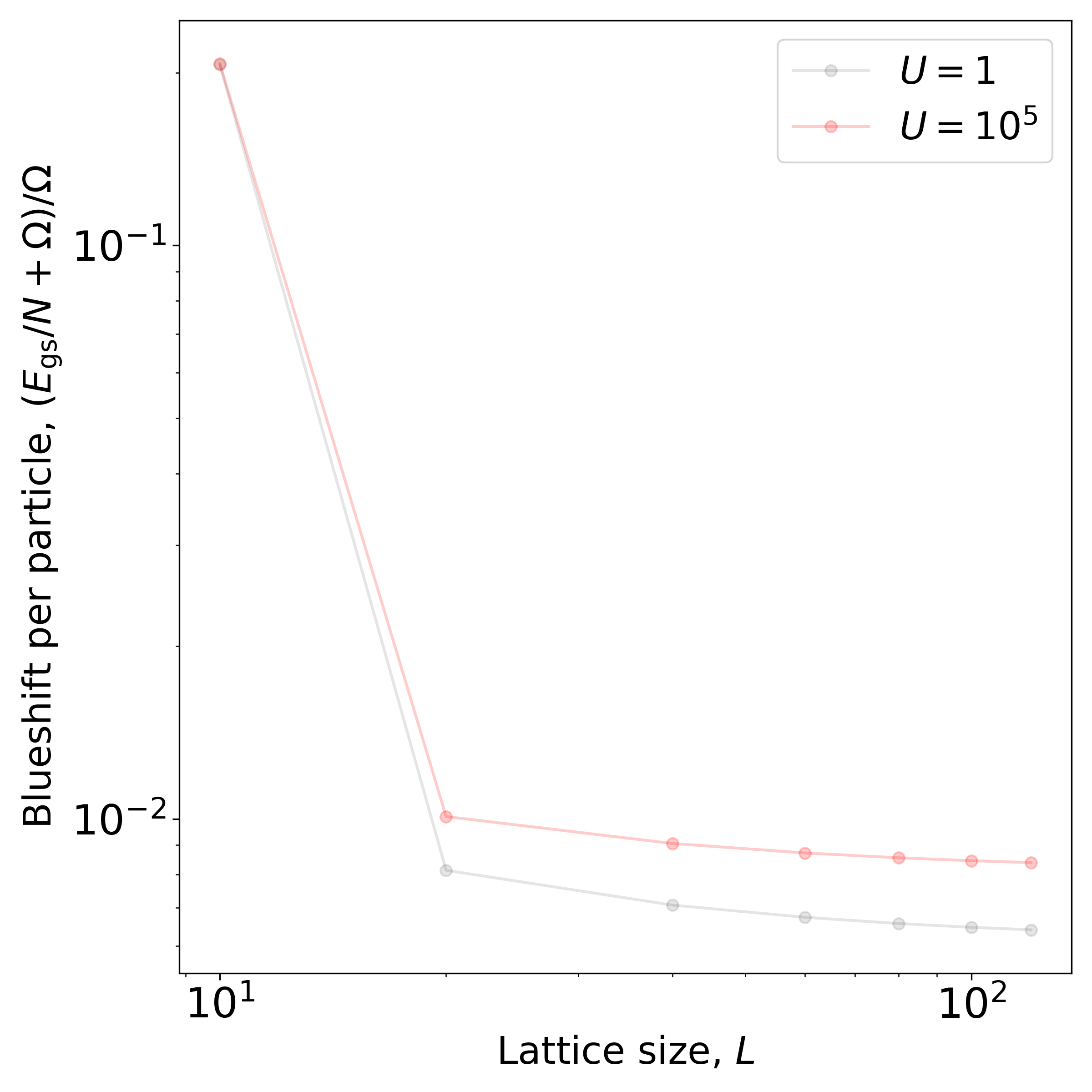}
    \caption{
    Blueshift per particle as a function of system size with open boundary conditions and at quarter filling.
    This is shown for different chain lengths from $L=10$ to $L=120$.
    Numerical parameters: $J/\Omega=0.1$, $U/\Omega = 1$.
    }
    \label{fig:finite_size_energies}
\end{figure}

We perform DMRG on this OBC system for lattice sizes $L \in \{ 10, 20, 40, 60, 80, 100, 120 \}$.
In Fig.~\ref{fig:finite_size_energies} it can be seen that the blueshift first decreases, as a function of $L$, and then quickly saturates.
Since this is shown on a logarithmic scale, it thus quickly \textit{converges} to the blueshift that one would obtain in the thermodynamic limit.
This is exemplarily shown both for the intermediate $U=1$ and a very large interaction strength, $U=10^5$.
But since the energy is expected to converge quickly, we also calculate the more sensitive correlations, that we also investigate in the main text.

\subsection{Correlations}

In the main text we show results for finite systems.
Figs.~\ref{fig:g2} and \ref{fig:Swq} are obtained for systems with open boundary conditions and a system size $L = 100$.
These systems are sufficiently large such that finite-size effects do not play an important role.
To show this numerically, we calculate the photonic correlation function, $g^{(2)}_\mathrm{ph}$, that was introduced in Eq.~\eqref{eq:photonic-correlator} in the main text.

\begin{figure}[t]
    \centering
\includegraphics[width=0.99\linewidth]{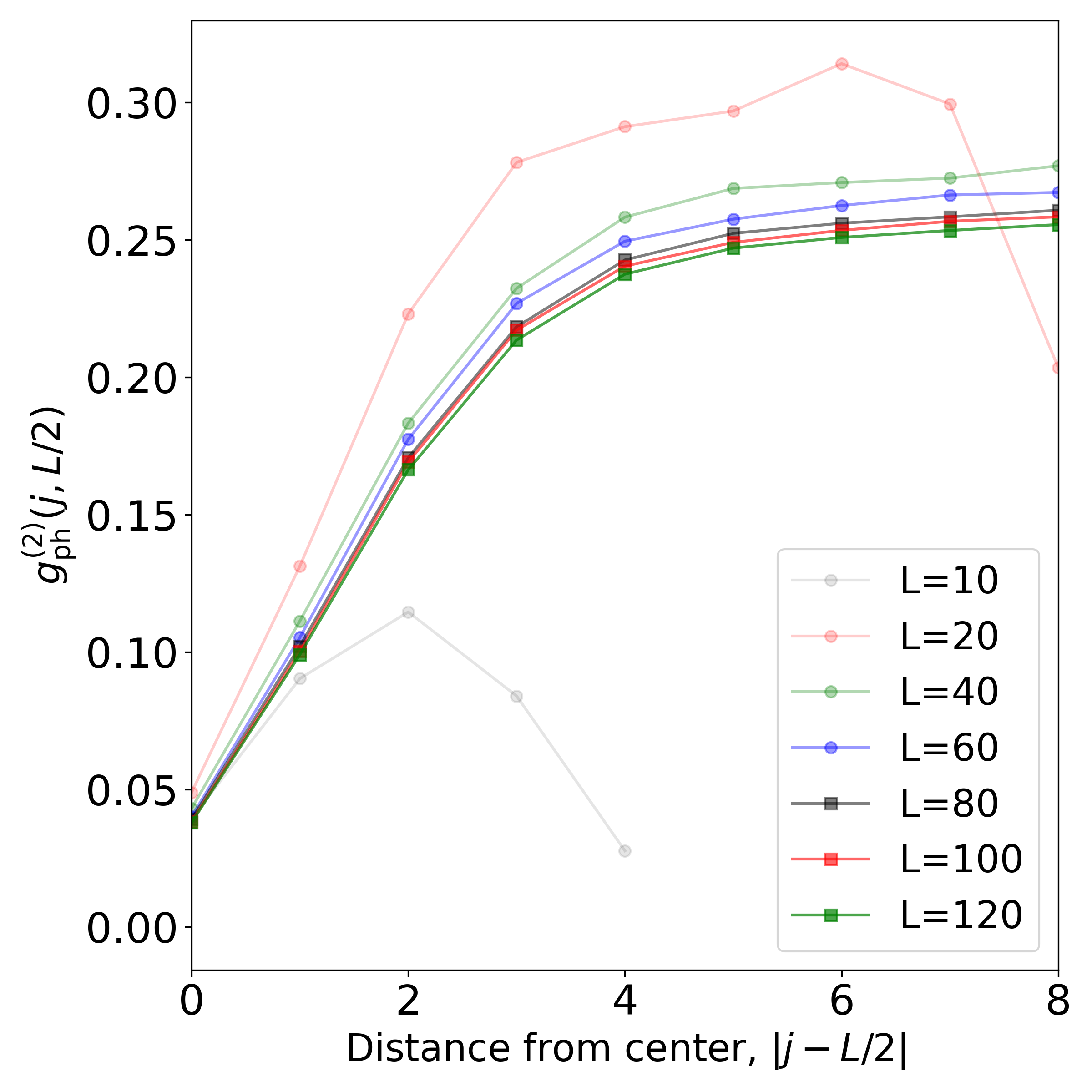}
    \caption{
    Finite-size scaling of photonic correlator as defined in Eq.~\eqref{eq:photonic-correlator}, evaluated $j$ sites away from the center of the finite chain.
    This is shown for different chain lengths from $L=10$ to $L=120$.
    Numerical parameters: $J/\Omega=0.1$, $U/\Omega = 1$.
    }
    \label{fig:finite_size}
\end{figure}

In Fig.~\ref{fig:finite_size} we evaluate $g^{(2)}_\mathrm{ph}(j,L/2)$ between a central site and a few sites away from the center.
It is apparent from these numerical results that the finite-size effects become less pronounced with growing system size.
While for a system with $L=10$, the edges of the system are quickly reached when successively going step by step away from the center, for the larger system we only probe their bulks.
As the systems grow in size, the correlators converge rather quickly as can be seen from a point-wise comparison of the different curves.
As expected, the excitonic correlators show the same tendency (results not shown).

\section{Photonic fraction \label{app:photonic-fraction}}

\begin{figure}[t]
    \centering
\includegraphics[width=0.99\linewidth]{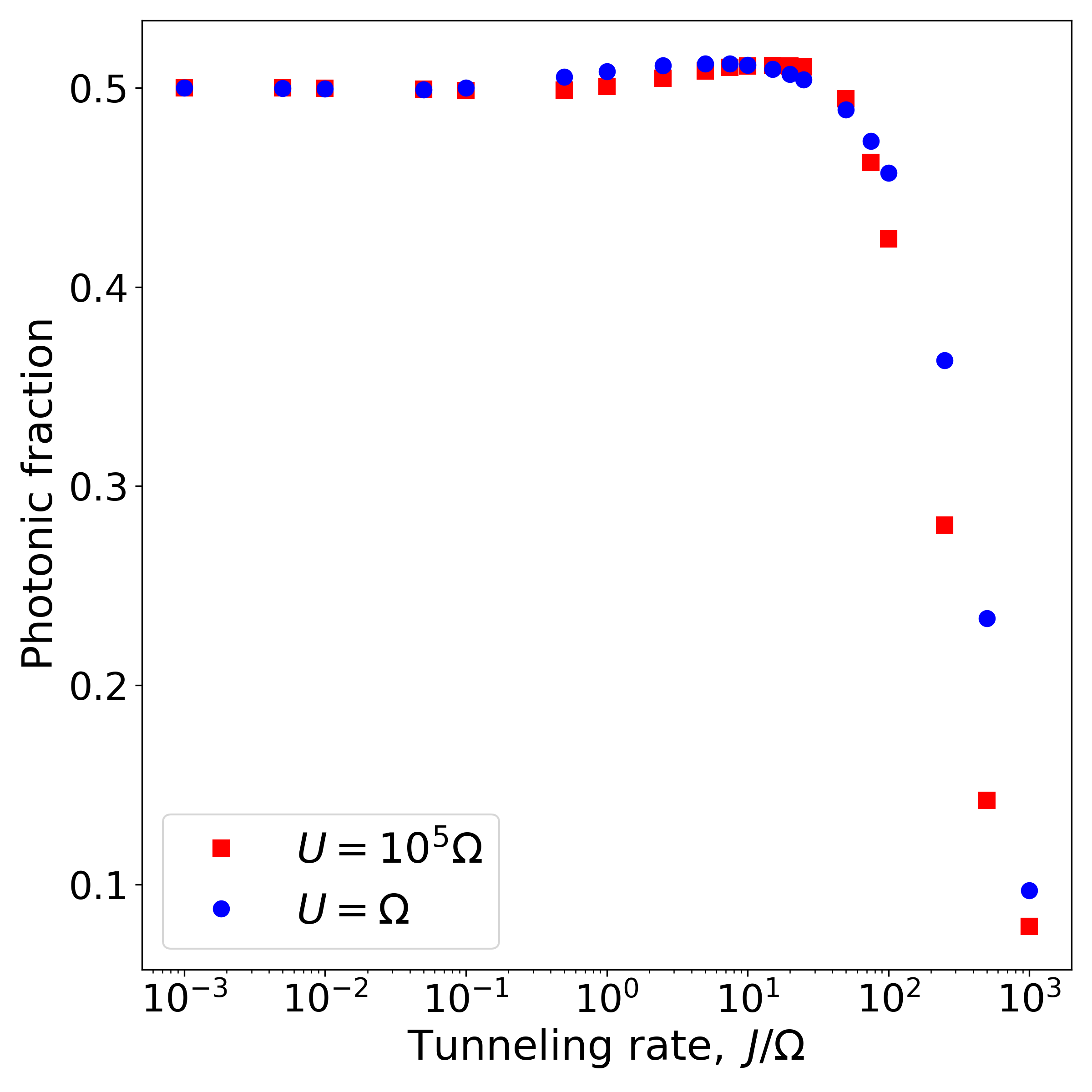}
    \caption{
    Photonic fraction as a function of tunneling rate, $J/\Omega$, for large (\textit{red squares}) and intermediate (\textit{blue circles}) on-site repulsion strengths $U$.
    Numerical parameters are $L = 60$, $N = 15$, and OBC are taken.
    }
    \label{fig:photonic_fraction}
\end{figure}

As discussed in the main text, the polaritons in our model are approximately \emph{half exciton}, \emph{half photon}, for large parts of parameter space.
However at really large $J/\Omega$ there are some corrections within PBC systems, and this picture completely fails for open boundary conditions.
For OBC, the photonic contribution disappears as we increase $J/\Omega$.
This is exemplarily shown in Fig.~\ref{fig:photonic_fraction} for a system at quarter filling, again for intermediate ($U/\Omega=1$) and large ($U/\Omega=10^5$) on-site interactions.
In both cases the matter component starts to dominate at tunneling rates that are much larger than $\Omega$.

\section{Von-Neumann entropy \label{app:von-neumann}}

\begin{figure}[t]
    \centering
\includegraphics[width=0.99\linewidth]{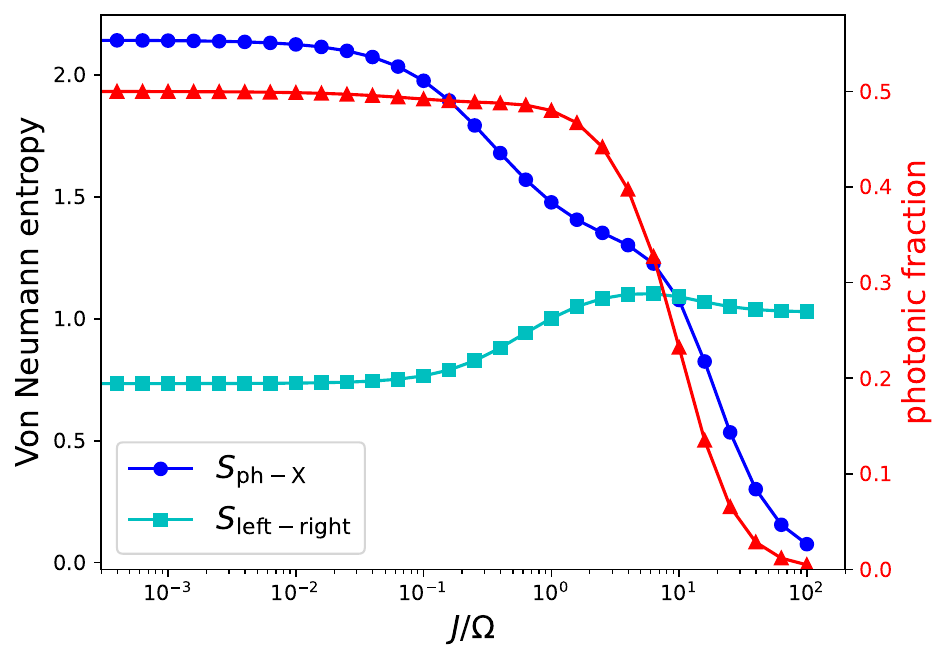}
    \caption{
   Exact diagonalization results for Von Neumann entropies (left axis) and photonic fraction (right axis, red triangles) as a function of $J/\Omega$. The Von Neumann entropy is computed by partitioning the system in two ways: cutting the rungs of the ladder to separate the photonic and excitonic degrees of freedom (blue circles); or cutting the left and right halves of the ladder (cyan squares).
   Simulation parameters:
   $L=8, N=3, U=\infty$ and OBC.
    }
    \label{fig:entanglement}
\end{figure}

It is also interesting to look at the light-matter entanglement as a function of $J/\Omega$.
A measure of the entanglement between two subsystems $A$ and $B$ is provided by the Von Neumann entropy
\begin{equation}
    S_AB = -{\rm Tr}_A[\rho_A\log\rho_A] 
    =
    -{\rm Tr}_B[\rho_B\log\rho_B]
\end{equation}
where $\rho_A = {\rm Tr}_B[|GS\rangle \langle GS|]$
is the reduced density matrix of subsystem $A$, and similarly for $\rho_B$.

In the following we will consider both the light-matter entanglement $S_{\rm X-ph}$ and the left-right one
$S_{\rm left-right}$.
Since the tensor network structure is not particularly 
convenient to compute 
$S_{\rm X-ph}$
(while $S_{\rm left-right}$ is straightforwardly obtained from the Schmidt decomposition of the central bond link), we use exact diagonalization on a small 
system 
to get a qualitative idea of the behavior of the Von Neumann entropy.

The exact diagonalization results are presented in 
Fig.~\ref{fig:entanglement} for a system with $L=8, N=3, U=\infty$ and open boundary conditions.
On the left vertical axis, the blue circles correspond to $S_{\rm X-ph}$ and the cyan squares to $S_{\rm left-right}$,
while the photonic fraction is displayed as red triangles (right vertical axis), for comparison with the DMRG results in a long system (Fig. \ref{fig:photonic_fraction}).
In particular, for very large $J$ there is no photonic component, because of the too large kinetic energy cost, the light-matter subsystems trivially decouple and the entropy is zero $S_{\rm X-ph}$. 
On the contrary, larger spatial correlations result in a larger $S_{\rm left-right}$.

\bibliography{bibliography}

\end{document}